\documentclass[3p,times]{elsarticle}

\usepackage{amsmath,amssymb,amsthm,mathtools}
\usepackage{algorithm}
\usepackage{algpseudocode}
\usepackage{booktabs}
\usepackage{hyperref}
\usepackage{microtype}
\usepackage{xspace}
\usepackage{xcolor}
\usepackage{tikz}

\biboptions{numbers,sort&compress}
\makeatletter
\def\ps@pprintTitle{%
  \let\@oddhead\@empty
  \let\@evenhead\@empty
  \let\@oddfoot\@empty
  \let\@evenfoot\@empty
}
\makeatother

\newtheorem{theorem}{Theorem}[section]
\newtheorem{lemma}[theorem]{Lemma}
\newtheorem{corollary}[theorem]{Corollary}
\newtheorem{proposition}[theorem]{Proposition}

\newtheorem{definition}[theorem]{Definition}
\newtheorem{problem}[theorem]{Problem}
\newtheorem{example}[theorem]{Example}

\newcommand{\MaxLL}{\textsc{Max-LL-DUP}\xspace}
\newcommand{\JISP}{\textsc{JISP}\xspace}
\newcommand{\JISPsub}[1]{\textsc{JISP}$_{#1}$\xspace}
\newcommand{\WJISP}{\textsc{WJISP}\xspace}
\newcommand{\LRS}{\textsc{LRS}\xspace}
\newcommand{\OPT}{\operatorname{OPT}}
\newcommand{\occ}{\operatorname{occ}}

\newcommand{\nxt}{\operatorname{next}}
\newcommand{\depth}{\operatorname{depth}}
\newcommand{\alphabet}{\operatorname{alph}}

\begin{document}

\begin{frontmatter}


\title{Maximum Letter-Duplicated and Longest Run Subsequences: Algorithms, Approximation, and Interval-Selection Connections\tnoteref{tnote1}}
\tnotetext[tnote1]{A preliminary version of part of this work appeared as ``Maximum Letter-Duplicated Subsequence (short paper)'' at the 25th Italian Conference on Theoretical Computer Science (ICTCS 2024)~\cite{DondiEtAl2024}.}

\author[bergamo]{Riccardo Dondi}
\ead{riccardo.dondi@unibg.it}
\author[bergamo]{Mohammad Mehdi Hosseinzadeh}
\ead{m.hosseinzadeh@unibg.it}
\author[bucharest]{Adrian Micl\u{a}u\c{s}}
\ead{adrian.miclaus@fmi.unibuc.ro}
\author[bucharest]{Alexandru Popa}
\ead{alexandru.popa@fmi.unibuc.ro}
\author[ichb]{Vlad Tutunaru}
\ead{vlad.tutunaru@ichb.ro}

\address[bergamo]{Universit\`a degli Studi di Bergamo, Bergamo, Italy}
\address[bucharest]{University of Bucharest, Bucharest, Romania}
\address[ichb]{International Computer High School of Bucharest, Bucharest, Romania}

\begin{abstract}
We study the \MaxLL problem. Given a string $S$, the goal is to find a letter-duplicated subsequence containing the maximum possible number of distinct symbols. A preliminary version of this work showed that the problem is solvable in linear time when every symbol occurs at most three times and is APX-hard when every symbol occurs at most four times. We develop a more complete algorithmic picture of the problem. Our main structural observation is that every feasible solution can be represented using intervals between consecutive occurrences of symbols, and the total number of relevant intervals is linear in the length of the input. This viewpoint gives a deterministic linear-time $2$-approximation for unrestricted \MaxLL and an exact reduction to the Job Interval Selection Problem. Consequently, for every constant $\varepsilon>0$, unrestricted \MaxLL admits a randomized $(e/(e-1)+\varepsilon)$-approximation, while the frequency-four case admits a randomized $(4/3+\varepsilon)$-approximation through an exact reduction to \JISPsub{2}. We also give an exact $O(2^p n)$-time algorithm, where $p$ is the number of symbols occurring more than three times. For $3$-\MaxLL range queries we reduce the problem to weighted-ancestor queries, obtaining linear space and $O(\log\log n)$ query time. We further study single-deletion sensitivity and show how all $n$ deletion answers can be computed in $O(2^p n+n\log n)$ total time. Finally, we consider the related Longest Run Subsequence problem. Both problems admit an interval-selection interpretation, but their objectives differ: one counts represented symbols whereas the other counts selected positions. Exploiting this connection, we give a value-preserving reduction to weighted Job Interval Selection and obtain a polynomial-time $2$-approximation without any restriction on symbol frequencies. To the best of our knowledge, this is the first constant-factor approximation for unrestricted Longest Run Subsequence.
\end{abstract}


\end{frontmatter}

\section{Introduction}
\label{sec:introduction}

\subsection{Motivation}

Gene, segmental, and whole-genome duplications are fundamental mechanisms in molecular evolution and are widely viewed as an important source of genomic novelty~\cite{Ohno1970,VanDePeer2004}. At the sequence level, a particularly simple operation is a tandem duplication, which copies a contiguous segment and inserts the copy immediately next to the original segment. In string notation this operation transforms $AXB$ into $AXXB$. Tandem duplications and their histories have therefore led to a variety of algorithmic problems on strings, including reconstruction and distance problems~\cite{LafondZhuZou2022}.

Subsequence formulations provide a complementary way to search for duplication signatures when the observed sequence has subsequently been affected by mutations, insertions, deletions, or assembly noise. For example, the Longest Run Subsequence problem was introduced in the context of homology-based genome scaffolding~\cite{SchrinnerEtAl2020,SchrinnerEtAl2021}. Lai et al.~\cite{LaiEtAl2022,LaiEtAl2024} introduced letter-duplicated subsequences as a simple model for recovering the skeleton of duplicated regions after local mutations. More recently, Lafond et al.~\cite{LafondEtAl2025} generalized this viewpoint from repeated letters to repeated subsequences.

A \emph{letter-duplicated subsequence} is, informally, a subsequence partitioned into consecutive blocks, where each block consists of at least two copies of one symbol and consecutive blocks use different symbols. Lai et al.~\cite{LaiEtAl2022,LaiEtAl2024} considered, among other variants, the requirement that every symbol of the input alphabet be represented. In many strings this requirement is too strong: different symbols may compete for overlapping portions of the input, so they cannot all be represented simultaneously. 
This motivates the maximization problem studied here, where the goal is instead to represent as many distinct symbols as possible.

\subsection{Informal problem definition}

The \MaxLL problem takes as input a string $S$ and asks for a letter-duplicated subsequence representing as many distinct symbols as possible. Since the objective counts symbols rather than subsequence length, once a symbol is represented we only need two of its occurrences. Thus one may think of a solution as choosing, for every selected symbol, a pair of equal positions, with the chosen pairs forming non-overlapping blocks from left to right.

A natural restriction is to bound the number of occurrences of every symbol. We write $d$-\MaxLL when each symbol occurs at most $d$ times. The preliminary conference paper~\cite{DondiEtAl2024} showed a sharp change between $d=3$ and $d=4$: $3$-\MaxLL is solvable in linear time, whereas $4$-\MaxLL is APX-hard. One goal of the present paper is to understand this threshold more fully; another is to study unrestricted instances through approximation, parameterization, and data structures.

\subsection{Previous and related work}

Lai et al.~\cite{LaiEtAl2022,LaiEtAl2024} initiated the study of letter-duplicated subsequences and proved hardness and polynomial-time results for variants requiring all alphabet symbols to be represented; the later journal version also studies a weighted formulation. Lafond et al.~\cite{LafondEtAl2025} subsequently considered the more general subsequence-duplicated subsequence model. Dondi, Hosseinzadeh, and Popa~\cite{DondiEtAl2024} introduced the maximization version studied here and proved the frequency-three linear-time algorithm and frequency-four APX-hardness result. We reproduce those two proofs in Sections~\ref{sec:frequency} and~\ref{sec:hardness}, using notation consistent with the journal version.

Our interval formulation relates \MaxLL to the Job Interval Selection Problem. Job Interval Selection and its weighted variants have been studied extensively; see Erlebach and Spieksma~\cite{ErlebachSpieksma2003} and Chuzhoy, Ostrovsky, and Rabani~\cite{ChuzhoyEtAl2006}. The latter give a randomized approximation ratio arbitrarily close to $e/(e-1)$ for general \JISP and a stronger guarantee when every job has only a bounded number of alternatives.

The range-query results use weighted-ancestor data structures and their connection with predecessor searching~\cite{BelazzouguiEtAl2021}. We also consider the related Longest Run Subsequence problem. LRS was introduced by Schrinner et al.~\cite{SchrinnerEtAl2020,SchrinnerEtAl2021} in genome scaffolding.
The \LRS problem is NP-hard, and it remains APX-hard even when every symbol occurs at most twice~\cite{DondiSikora2021}.
The problem was further studied from the viewpoint of parameterized complexity by Dondi and Sikora~\cite{DondiSikora2021}. Approximation algorithms for bounded-occurrence instances were developed by Asahiro et al.~\cite{AsahiroEtAl2023,AsahiroEtAl2025}. Their guarantees depend on the maximum number $k$ of occurrences of a symbol. Our reduction to weighted Job Interval Selection yields a factor $2$ independent of $k$. To the best of our knowledge, this is the first constant-factor approximation for unrestricted LRS.

\subsection{Our results}
\label{subsec:our-results}

Section~\ref{sec:framework} formally defines both problems and collects the notation and structural preliminaries used throughout the paper. If the occurrences of a symbol $a$ are $p^a_1<\cdots<p^a_{t_a}$, we show that it is enough to consider the consecutive-occurrence intervals $[p^a_i,p^a_{i+1}]$. Hence a string of length $n$ yields at most $n-1$ relevant intervals.

Section~\ref{sec:frequency} presents the linear-time dynamic-programming algorithm for $3$-\MaxLL from the ICTCS paper and builds on it to answer substring queries through weighted ancestors. Section~\ref{sec:hardness} reproduces the ICTCS cubic-Maximum-Independent-Set reduction for $4$-\MaxLL, includes a figure illustrating its gadgets, and derives the exact optimum relation and an explicit approximation lower bound. In particular, hardness holds when every symbol occurs either twice or four times.

Section~\ref{sec:approximation} contains our approximation results. A direct greedy algorithm is a deterministic $2$-approximation and runs in $O(n)$ time. We then show that \MaxLL is exactly an instance of \JISP, yielding a randomized $(e/(e-1)+\varepsilon)$-approximation from Chuzhoy et al.~\cite{ChuzhoyEtAl2006}. For $4$-\MaxLL, the special three-interval structure of a four-occurring symbol gives an exact reduction to \JISPsub{2}, and hence a randomized $(4/3+\varepsilon)$-approximation. Combined with the conference hardness result, this yields a complete frequency dichotomy: for every fixed $d\le3$, $d$-\MaxLL is solvable in linear time, while for every fixed $d\ge4$ it is APX-complete.

Section~\ref{sec:fpt} parameterizes the problem by $p=|\{a\in\Sigma:\occ_S(a)>3\}|$, the number of symbols violating the frequency-three tractability threshold, and gives an exact $O(2^pn)$-time dynamic program.

Section~\ref{sec:sensitivity} considers edits and independent single deletions, and gives an offline algorithm that computes all $n$ deletion answers in $O(2^pn+n\log n)$ total time.

Finally, Section~\ref{sec:lrs} develops the connection with Longest Run Subsequence, defined in Section~\ref{sec:framework}. Existing approximation ratios are expressed in terms of the maximum occurrence bound $k$~\cite{AsahiroEtAl2023,AsahiroEtAl2025}; in particular, the most recent bounds are $(k+2)/3$ for $k\in\{2,3\}$ and $2(k+1)/5$ for $k\ge4$. We show that Longest Run Subsequence reduces exactly to weighted Job Interval Selection on one machine, yielding a polynomial-time $2$-approximation by known scheduling results~\cite{ErlebachSpieksma2003,BarNoyEtAl2001,BermanDasGupta2000}. To the best of our knowledge, this is the first constant-factor approximation for unrestricted LRS, with no bound on symbol frequencies.

\section{Preliminaries}
\label{sec:framework}

\subsection{String notation and formal problems}

Let $S=s_1s_2\cdots s_n$ be a string over a finite alphabet
$\Sigma$. For $a\in\Sigma$, let $\occ_S(a)$ be the number of
occurrences of $a$ in $S$. When the string is clear, we write
$\occ(a)$. A string $T=t_1\cdots t_\ell$ is a subsequence of
$S$, denoted $T\preceq S$, if there are indices
$1\le i_1<\cdots<i_\ell\le n$ such that $t_j=s_{i_j}$ for
every $j$.

For a string $T$, let $\alphabet(T)$ be the set of symbols
occurring in $T$. For $1\le \ell\le r\le n$, the notation
$S[\ell..r]$ denotes the contiguous substring
$s_\ell s_{\ell+1}\cdots s_r$. We also allow the empty prefix
$S[1..0]$ and the empty suffix $S[n+1..n]$. For a position
$x$, we write $S-x$ for the string obtained by deleting the
occurrence at position $x$. A \emph{run} of a string is a
maximal nonempty block of identical symbols.

\begin{definition}[Letter-duplicated subsequence]
A subsequence $T\preceq S$ is \emph{letter-duplicated} if it
can be written as $T=a_1^{\ell_1}a_2^{\ell_2}\cdots
a_r^{\ell_r}$, where $\ell_i\ge 2$ for every $i$ and
$a_i\ne a_{i+1}$ for every $i<r$.
\end{definition}

\begin{problem}[Maximum Letter-Duplicated Subsequence (\MaxLL)]
Given a string $S$, find a letter-duplicated subsequence
$T\preceq S$ maximizing $|\alphabet(T)|$.
\end{problem}

For an integer $d\ge 1$, $d$-\MaxLL denotes the restriction
in which $\occ(a)\le d$ for every $a\in\Sigma$.

\begin{definition}[Run subsequence]
A subsequence $T\preceq S$ is a \emph{run subsequence} if
every symbol occurring in $T$ occurs in exactly one run.
Equivalently, $T=a_1^{\ell_1}\cdots a_t^{\ell_t}$, where
$a_1,\ldots,a_t$ are pairwise distinct symbols and
$\ell_i\ge 1$ for every $i$.
\end{definition}

\begin{problem}[Longest Run Subsequence (\LRS)]
Given a string $S$, find a run subsequence $T\preceq S$
maximizing its length $|T|$.
\end{problem}

Similarly, for an integer $k\ge 1$, $k$-\LRS denotes the
restriction in which $\occ(a)\le k$ for every $a\in\Sigma$.

The two problems are closely related but differ in both their
feasibility constraints and their objective functions. In
\MaxLL, every run must have length at least two, and the
objective is to maximize the number of distinct symbols in
the subsequence. In contrast, \LRS allows runs of length one
but requires each symbol to occur in at most one run, and
the objective is to maximize the total length of the
subsequence. Nevertheless, since the objective of \MaxLL
counts distinct symbols, there always exists an optimal
solution in which each symbol is represented by a single
run. This common property is the basis of the interval
formulations developed in this paper.

\begin{example}
\label{ex:two-problems}
Let $S=\mathtt{abaaabcc}$. The subsequence $\mathtt{aacc}$
is letter-duplicated and contains two distinct symbols.
No feasible solution of \MaxLL can contain all three
symbols: selecting both occurrences of $\mathtt{b}$
leaves only one occurrence of $\mathtt{a}$ outside the
interval spanned by the two occurrences of $\mathtt{b}$.
Hence $\OPT_{\MaxLL}(S)=2$.

For \LRS, the subsequence $\mathtt{aaaabcc}$ is a feasible
run subsequence of length seven, obtained by deleting the
first occurrence of $\mathtt{b}$. Since $S$ itself is not
a run subsequence, no feasible solution has length eight.
Consequently, $\OPT_{\LRS}(S)=7$.
\end{example}

\subsection{The occurrence-interval framework}

For every symbol $a$, write its occurrence positions as $p^a_1<p^a_2<\cdots<p^a_{t_a}$. We associate with $a$ the consecutive-occurrence intervals $I^a_i=[p^a_i,p^a_{i+1}]$ for $1\le i<t_a$.

\begin{lemma}[Consecutive-occurrence normal form]
\label{lem:consecutive}
For every feasible solution $T$ of \MaxLL, there exists a feasible solution $T'$ with $\alphabet(T')=\alphabet(T)$ such that every symbol $a\in\alphabet(T')$ is represented by two consecutive occurrences $p^a_i,p^a_{i+1}$ in $S$.
\end{lemma}

\begin{proof}
Because the objective counts distinct symbols, for each symbol used by $T$ we may retain a single run and two occurrences from that run. The intervals spanned by the retained pairs are pairwise disjoint. Suppose the retained pair for $a$ is $p^a_i,p^a_j$ with $j>i+1$. 
The earlier endpoint of the pair may be retained, while the later one is moved left to the next occurrence of the same symbol; the order of all other chosen pairs is unchanged.
Replacing it by $p^a_i,p^a_{i+1}$ replaces $[p^a_i,p^a_j]$ by a contained interval and hence cannot create a new intersection with any other retained interval. Applying this replacement independently to every non-consecutive pair preserves feasibility and the represented symbol set.
\end{proof}

Let $\mathcal I(S)=\{I^a_i:a\in\Sigma,\ 1\le i<t_a\}$. The normal form immediately gives $|\mathcal I(S)|=\sum_{a:\occ(a)>0}(t_a-1)=n-|\{a:\occ(a)>0\}|\le n-1$.

\begin{proposition}
\label{prop:interval-characterization}
A set $A\subseteq\Sigma$ is representable by a feasible \MaxLL solution if and only if there is a pairwise disjoint family $\mathcal F\subseteq\mathcal I(S)$ containing exactly one interval associated with every symbol of $A$ and no interval associated with a symbol outside $A$.
\end{proposition}

\begin{proof}
The forward implication is Lemma~\ref{lem:consecutive}. Conversely, order the intervals of $\mathcal F$ from left to right and select their two endpoints. Since the intervals are pairwise disjoint and every symbol contributes at most one interval, these endpoints form a letter-duplicated subsequence containing exactly the symbols of $A$.
\end{proof}

When endpoint conventions matter, every closed interval $[u,v]$ may equivalently be replaced by the half-open interval $[2u,2v+1)$. This preserves the intended fact that two candidate intervals sharing an occurrence overlap.

\subsection{Scheduling terminology and parameters}

In the Job Interval Selection Problem (\JISP), each job has a finite set of possible intervals. A feasible schedule chooses at most one interval from every job, and the selected intervals must be pairwise non-overlapping. The objective is to maximize the number of scheduled jobs. In the weighted version (\WJISP), intervals carry positive weights and the objective is to maximize total selected weight~\cite{ErlebachSpieksma2003,ChuzhoyEtAl2006}. The reduction for \MaxLL chooses only intervals between consecutive occurrences and assigns value one per selected symbol; the reduction for \LRS will instead use weighted intervals spanning arbitrary first and last occurrences of a run. This is the structural connection between the two problems.

For the parameterized results we use $D(S)=\{a\in\Sigma:\occ_S(a)>3\}$ and $p=|D(S)|$. Thus $p$ counts the symbols that lie outside the frequency-three tractable regime. For range queries the string $S$ is fixed, and a query $[\ell,r]$ asks for the optimum value on the substring $S[\ell..r]$.

\section{Polynomial-time algorithms and substring queries for $3$-\MaxLL}
\label{sec:frequency}

We first recall the linear-time algorithm for frequency-three inputs from the preliminary ICTCS paper~\cite{DondiEtAl2024}, including its dynamic-programming proof. We then develop a data structure for answering many substring queries without rerunning the algorithm for every query. Both results exploit the fact that, with at most three occurrences per symbol, intervals associated with the same symbol always overlap.

\subsection{A linear-time algorithm for $3$-\MaxLL}

For $j\in\{0,\ldots,n\}$, let $D[j]$ denote the optimum value of $3$-\MaxLL on the prefix $S[1..j]$. We set $D[0]=D[1]=0$. For $j\ge2$, let $a=s_j$ and let $r_a(j)$ be the rightmost occurrence of $a$ in $S[1..j-1]$, when such an occurrence exists. If $r_a(j)$ is undefined, then $D[j]=D[j-1]$; otherwise,
\begin{equation}
\label{eq:three-dp}
D[j]=\max\{D[j-1],\,D[r_a(j)-1]+1\}.
\end{equation}

\begin{lemma}
\label{lem:three-dp-correct}
For every $j\in\{0,\ldots,n\}$, $D[j]$ equals the optimum value of $3$-\MaxLL on $S[1..j]$.
\end{lemma}

\begin{proof}
We proceed by induction on $j$. The claim is immediate for $j=0,1$. Assume it holds for every smaller prefix and consider an optimum solution $T$ for $S[1..j]$.

If $T$ does not use position $j$, then $T$ is feasible for $S[1..j-1]$, so its value is at most $D[j-1]$. Suppose instead that $T$ uses $s_j=a$. Since $a$ must occur at least twice in its run, $T$ uses an earlier occurrence of $a$. We may choose the rightmost such occurrence $r_a(j)$, because replacing an earlier copy by $s_{r_a(j)}$ can only shorten the final block. Removing the pair representing $a$ leaves a feasible solution contained in $S[1..r_a(j)-1]$. Hence the value of $T$ is at most $D[r_a(j)-1]+1$.

Conversely, an optimum solution for $S[1..j-1]$ is also feasible for $S[1..j]$. If $r_a(j)$ exists, take an optimum solution for $S[1..r_a(j)-1]$ and append the pair $s_{r_a(j)}s_j$. Because every symbol occurs at most three times in the entire input, the prefix $S[1..r_a(j)-1]$ contains at most one occurrence of $a$. Thus its optimum solution cannot already represent $a$, and appending the pair produces a feasible solution of value $D[r_a(j)-1]+1$. Equation~\eqref{eq:three-dp} therefore gives exactly the optimum.
\end{proof}

\begin{theorem}[Dondi et al.~\cite{DondiEtAl2024}]
\label{thm:three-linear}
$3$-\MaxLL can be solved in $O(n)$ time.
\end{theorem}

\begin{proof}
By Lemma~\ref{lem:three-dp-correct}, it suffices to evaluate~\eqref{eq:three-dp}. Scan $S$ from left to right while storing the most recent occurrence of each symbol. The value $r_a(j)$ is therefore available in constant time, and each entry $D[j]$ is computed in constant time from earlier entries. The table has $n+1$ entries, so the total running time is $O(n)$. Standard predecessor pointers recover an optimum subsequence within the same asymptotic bound.
\end{proof}

The occurrence-interval framework gives an alternative view of the same threshold: when a symbol occurs at most three times it has at most two candidate intervals, and if two exist they intersect at the middle occurrence. Hence pairwise disjointness automatically enforces the at-most-one-interval-per-symbol condition.

\subsection{Substring queries by weighted ancestors}
\label{sec:range}
The dynamic program above solves a single instance in linear time. In applications where many subintervals of the same input sequence are queried, independently running the algorithm on each substring would be unnecessarily expensive. We therefore preprocess a fixed string $S=s_1\cdots s_n$ in which every symbol occurs at most three times so that a query $[\ell,r]$ returns the optimum value on $S[\ell..r]$. The key is to encode the choices of the earliest-finishing-time greedy interval-scheduling algorithm, which is equivalent to the frequency-three dynamic program by the occurrence-interval characterization.

For every position $i\in\{1,\ldots,n\}$, let $f(i)$ be the next occurrence of $s_i$, i.e., $f(i)=\min\{j>i:s_j=s_i\}$, with $f(i)=n+1$ if no such position exists. For $i\in\{0,\ldots,n\}$, define $\nxt(i)=\min_{j>i}f(j)$, with $\nxt(i)=n+1$ if no candidate interval starts after $i$. Set $\nxt(n+1)=n+1$. The values $\nxt(i)$ satisfy $\nxt(i)>i$ for $i\le n$ and can be computed in $O(n)$ time by a right-to-left scan using $\nxt(i)=\min\{f(i+1),\nxt(i+1)\}$.

Here the greedy algorithm is the earliest-finishing-time procedure for unweighted interval scheduling, which is optimal in the frequency-three case. It differs from the general $2$-approximation algorithm of Section~\ref{sec:approximation}, which additionally checks whether a symbol was previously selected.

\begin{lemma}
\label{lem:next-greedy}
If the right endpoint of the previously selected interval is $i$, then, whenever $\nxt(i)\le n$, the next interval chosen by the earliest-finishing-time greedy algorithm ends at $\nxt(i)$.
\end{lemma}

\begin{proof}
A candidate interval $[j,f(j)]$ is compatible with the previous choice exactly when $j>i$. The greedy algorithm chooses a compatible interval with minimum right endpoint, which is $\min_{j>i}f(j)=\nxt(i)$.
\end{proof}

For $q\ge0$, write $\nxt^q$ for $q$ successive applications of the function $\nxt$, with $\nxt^0(i)=i$. Thus a query $[\ell,r]$ asks for the largest integer $q$ such that $\nxt^q(\ell-1)\le r$.

Create a rooted tree on vertices $\{0,1,\ldots,n+1\}$ by setting the parent of $i\le n$ to $\nxt(i)$ and using $n+1$ as the root. Since $\nxt(i)>i$, this is a rooted tree. Give node $i$ weight $w(i)=n+1-i$; weights strictly increase from the root towards the leaves. For a query $[\ell,r]$, let $u=\ell-1$ and let $z=n+1-r$. 

The meaning of this ancestor query can be seen directly along the parent chain $u=u_0,u_1,u_2,\ldots$, where $u_{j+1}=\nxt(u_j)$. This chain is strictly increasing until it reaches the sentinel $n+1$. By Lemma~\ref{lem:next-greedy}, precisely the terms $u_1,\ldots,u_q$ with $u_j\le r$ are right endpoints selected by the greedy algorithm on $S[\ell..r]$. Since the node weights are $w(i)=n+1-i$, the condition $u_j\le r$ is equivalent to $w(u_j)\ge n+1-r=z$. Hence the last accepted term of the parent chain is exactly the highest ancestor $v$ of $u$ whose weight is at least $z$. Each parent edge represents one greedy selection, so the number of selected intervals is the number of parent edges from $u$ to $v$, namely $\depth(u)-\depth(v)$.

\begin{example}\label{ex:range}
Let $S=\mathtt{abbaacc}$ and consider the query $[2,5]$, for which $S[2..5]=\mathtt{bbaa}$. We have $\nxt(1)=3$ (the interval for $\mathtt b$ is $[2,3]$), and $\nxt(3)=5$ (the interval for $\mathtt a$ is $[4,5]$). The next value is the sentinel $n+1=8$. Thus exactly two transitions have a right endpoint at most $r=5$. In the parent tree, the accepted ancestor is $v=5$ of $u=1$, and $\depth(1)-\depth(5)=2$, the optimum value on the query substring.
\end{example}

Thus the weighted-ancestor structure returns $v$ and the answer is $\depth(u)-\depth(v)$.

The general weighted-ancestor problem on an $O(n)$-node tree with integer weights from an $O(n)$ universe can be supported in linear space with $O(\log\log n)$ query time via predecessor-search data structures~\cite{BelazzouguiEtAl2021}.

\begin{theorem}
\label{thm:range}
$3$-\MaxLL can be preprocessed into an $O(n)$-space data structure that answers every substring query in $O(\log\log n)$ time. A straightforward implementation has $O(n\log\log n)$ preprocessing time; standard static predecessor constructions can be used to obtain linear or expected-linear construction under the usual word-RAM assumptions.
\end{theorem}

\begin{proof}
Compute $f$, $\nxt$, the rooted tree, and all node depths. Store a weighted-ancestor structure for the tree. By Lemma~\ref{lem:next-greedy}, the number of selected intervals in a query is exactly the number of parent edges from $u=\ell-1$ to the highest ancestor whose index is at most $r$, and this number equals the depth difference returned above. The stated space and query bounds follow from the weighted-ancestor data structure.
\end{proof}

\section{Hardness of $4$-\MaxLL}
\label{sec:hardness}
\subsection{The cubic independent-set reduction}

We next turn to the first frequency bound at which the problem becomes APX-hard. We reproduce the reduction of~\cite{DondiEtAl2024}, with the details needed for the later approximation lower bound. The source problem is Maximum Independent Set on cubic graphs, which is APX-complete~\cite{AlimontiKann2000}. Let $G=(V,E)$ be a cubic graph with $V=\{v_1,\ldots,v_N\}$.

For each vertex $v_i$, introduce symbols $a_i$ and $b_i$. For every edge $\{v_i,v_j\}\in E$, introduce two oriented symbols $e_{i,j}$ and $e_{j,i}$. Thus $\Sigma=\{a_i,b_i:v_i\in V\}\cup\{e_{i,j},e_{j,i}:\{v_i,v_j\}\in E\}$.

For a vertex $v_i$ with neighbors $v_j,v_z,v_w$, define $S_1(v_i)=a_i b_i a_i b_i$ and $S_2(v_i)=a_i e_{i,j}e_{i,j}e_{i,z}e_{i,z}e_{i,w}e_{i,w}a_i$. For every edge $\{v_i,v_j\}\in E$ with $i<j$, define $S(i,j)=e_{i,j}e_{j,i}e_{i,j}e_{j,i}$. The string $S(G)$ is obtained by concatenating all blocks $S_1(v_i)$, then all blocks $S_2(v_i)$, and finally all edge blocks $S(i,j)$, in any fixed order within each group. We call $a_i,b_i$ and the three oriented edge symbols corresponding to edges incident with $v_i$ the five symbols \emph{related to $v_i$}.
An example of the reduction is given in Figure~\ref{fig:cubic-reduction}.

\begin{figure}[t]
\centering
\begin{minipage}{0.35\linewidth}
\centering
\begin{tikzpicture}[scale=0.91, every node/.style={circle,draw,inner sep=2pt,minimum size=0.55cm}]
\node[fill=gray!25] (a) at (0,2) {$v_1$};
\node (b) at (2,2) {$v_2$};
\node (c) at (0,0) {$v_3$};
\node (d) at (2,0) {$v_4$};
\draw (a)--(b) (a)--(c) (a)--(d) (b)--(c) (b)--(d) (c)--(d);
\end{tikzpicture}
\end{minipage}\hfill
\begin{minipage}{0.62\linewidth}
\small
$S_1(v_1)=a_1b_1a_1b_1$\\[3pt]
$S_2(v_1)=a_1e_{1,2}e_{1,2}e_{1,3}e_{1,3}e_{1,4}e_{1,4}a_1$\\[3pt]
$S(1,2)=e_{1,2}e_{2,1}e_{1,2}e_{2,1}$
\end{minipage}
\caption{Illustration of the reduction on the cubic graph $K_4$. A vertex contributes two local blocks, and an edge contributes one interleaved edge block. The shaded vertex $v_1$ is an independent set of size one; its five associated symbols can be represented, while each other vertex contributes four symbols, giving the value $4\cdot4+1=17$.}
\label{fig:cubic-reduction}
\end{figure}

\begin{lemma}
\label{lem:hard-frequency}
Every symbol in $S(G)$ occurs at most four times. More precisely, every $b_i$ occurs twice, while every $a_i$ and every oriented edge symbol $e_{i,j}$ occurs four times.
\end{lemma}

\begin{proof}
The symbol $b_i$ appears only in $S_1(v_i)$, where it occurs twice. The symbol $a_i$ occurs twice in $S_1(v_i)$ and twice in $S_2(v_i)$. Finally, $e_{i,j}$ occurs twice in $S_2(v_i)$ and twice in the edge block $S(i,j)$. No other occurrences are introduced.
\end{proof}

\begin{lemma}
\label{lem:hard-forward}
If $G$ has an independent set $I$, then $S(G)$ has a feasible $4$-\MaxLL solution of value $5|I|+4(N-|I|)=4N+|I|$.
\end{lemma}

\begin{proof}
For each $v_i\in I$, select the pair $b_i b_i$ inside $S_1(v_i)$ and the pair $a_i a_i$ given by the two endpoint occurrences of $S_2(v_i)$. For every edge incident with $v_i$, select the pair of the corresponding oriented edge symbol $e_{i,j}$ inside the edge block $S(i,j)$. Thus all five symbols related to $v_i$ are represented.

For each $v_i\notin I$, select the pair $a_i a_i$ inside $S_1(v_i)$ and, inside $S_2(v_i)$, select the three pairs of the oriented edge symbols incident with $v_i$. Thus four symbols related to $v_i$ are represented.

These choices are feasible. Inside an edge block $S(i,j)$, the pairs for $e_{i,j}$ and $e_{j,i}$ are incompatible; both would be selected there only if $v_i,v_j\in I$, which is impossible because $I$ is independent. All remaining selected pairs occur in disjoint concatenation blocks. Hence the resulting solution represents $5|I|+4(N-|I|)$ symbols.
\end{proof}

\begin{lemma}
\label{lem:hard-backward}
Suppose $S(G)$ has a feasible solution of value $4N+q$. Then one can compute in polynomial time an independent set $I$ of $G$ with $|I|\ge q$.
\end{lemma}

\begin{proof}
The five symbols related to distinct vertices form disjoint sets. Since at most five symbols related to any one vertex can be represented, a solution of value $4N+q$ must represent all five related symbols for at least $q$ vertices. Let $I$ be the set of such vertices; then $|I|\ge q$.

Fix $v_i\in I$. Since $b_i$ occurs only in $S_1(v_i)$, the solution must use the pair $b_i b_i$ in that block. Consequently, it cannot represent $a_i$ by the two occurrences in $S_1(v_i)$, because in the pattern $a_i b_i a_i b_i$ the two pairs are incompatible. Nor can a pair for $a_i$ use one occurrence in $S_1(v_i)$ and a later occurrence: such a pair would contain the selected pair $b_i b_i$ and would again be incompatible. Hence $a_i$ must be represented by the two endpoint occurrences of $S_2(v_i)$.

It follows that none of the three oriented edge symbols related to $v_i$ can be represented by its pair inside $S_2(v_i)$, because each such pair lies strictly between the two selected occurrences of $a_i$. Therefore, for every neighbor $v_j$ of $v_i$, the pair for $e_{i,j}$ must be selected in the edge block $S(i,j)$.

Now suppose that two vertices $v_i,v_j\in I$ are adjacent. In $S(i,j)=e_{i,j}e_{j,i}e_{i,j}e_{j,i}$, the pairs for $e_{i,j}$ and $e_{j,i}$ are incompatible, yet the preceding argument requires both. This is impossible. Thus $I$ is an independent set of size at least $q$.
\end{proof}

\begin{theorem}[Dondi et al.~\cite{DondiEtAl2024}]
\label{thm:four-hard}
$4$-\MaxLL is APX-hard.
\end{theorem}

\begin{proof}
The preceding lemmas give an $L$-reduction, in the sense of Papadimitriou and Yannakakis~\cite{PapadimitriouYannakakis1991}, from cubic Maximum Independent Set. First, a baseline solution of value $4N$ is obtained from Lemma~\ref{lem:hard-forward} with $I=\emptyset$. If an optimum target solution has value $4N+q$, Lemma~\ref{lem:hard-backward} gives $q\le\alpha(G)$, while Lemma~\ref{lem:hard-forward} applied to a maximum independent set gives the reverse inequality. Hence $\OPT_{\MaxLL}(S(G))=4N+\alpha(G)$.

Every cubic graph has an independent set of size at least $N/4$, so $N\le4\alpha(G)$ and $\OPT_{\MaxLL}(S(G))\le17\alpha(G)$. For any feasible target solution, keep the better of it and the baseline solution and write its value as $4N+q$. Lemma~\ref{lem:hard-backward} produces an independent set $I$ with $|I|\ge q$, and therefore $\alpha(G)-|I|\le\alpha(G)-q=\OPT_{\MaxLL}(S(G))-(4N+q)$. Thus the two $L$-reduction constants may be taken as $17$ and $1$. Since Maximum Independent Set is APX-hard on cubic graphs~\cite{AlimontiKann2000}, $4$-\MaxLL is APX-hard.
\end{proof}

\begin{corollary}
\label{cor:two-four-hard}
$4$-\MaxLL remains APX-hard when every symbol occurs either exactly twice or exactly four times.
\end{corollary}

\begin{proof}
This follows directly from Lemma~\ref{lem:hard-frequency}.
\end{proof}

\subsection{Exact correspondence and explicit inapproximability}
The proof of Theorem~\ref{thm:four-hard} in fact establishes an exact optimum relation. We record it separately to make the quantitative hardness consequence explicit.

\begin{lemma}
\label{lem:exact-reduction}
The reduction satisfies $\OPT_{4\text{-}\mathrm{Max\text{-}LL\text{-}DUP}}(S(G))=4N+\alpha(G)$, where $\alpha(G)$ is the maximum independent-set size of $G$.
\end{lemma}

\begin{proof}
Lemma~\ref{lem:hard-forward} gives an output of value $4N+\alpha(G)$ from a maximum independent set. Conversely, Lemma~\ref{lem:hard-backward} applied to an optimum solution of value $4N+q$ yields $q\le\alpha(G)$. These two inequalities give the equality.
\end{proof}

Chleb\'ik and Chleb\'ikov\'a~\cite{ChlebikChlebikova2006} show that Maximum Independent Set on graphs of maximum degree three, even on three-regular graphs, is NP-hard to approximate within $95/94$.

\begin{theorem}
\label{thm:explicit-hardness}
For every $\varepsilon>0$, it is NP-hard to approximate $4$-\MaxLL within $1615/1614-\varepsilon$, even when every symbol occurs either exactly twice or exactly four times.
\end{theorem}

\begin{proof}
Assume that a $\rho$-approximation exists for these \MaxLL instances. We may compare its output with the easy baseline solution of value $4N$, so write the returned value as $4N+q$ with $q\ge 0$. By Lemma~\ref{lem:hard-backward}, we obtain an independent set of size at least $q$. Lemma~\ref{lem:exact-reduction} gives $4N+q\ge(4N+\alpha(G))/\rho$, hence $q\ge(4N+\alpha(G))/\rho-4N$. Every cubic graph satisfies $\alpha(G)\ge N/4$, so $N\le 4\alpha(G)$ and therefore $q\ge\alpha(G)(17-16\rho)/\rho$. Thus the induced approximation ratio for cubic Maximum Independent Set is at most $\rho/(17-16\rho)$. Equality with $95/94$ occurs at $\rho=1615/1614$. Any smaller constant would contradict the hardness bound of~\cite{ChlebikChlebikova2006}. Corollary~\ref{cor:two-four-hard} gives the stated occurrence restriction.
\end{proof}

\section{Approximation algorithms}
\label{sec:approximation}

\subsection{A deterministic linear-time $2$-approximation}

Consider the intervals of $\mathcal I(S)$ in nondecreasing order of right endpoint. The algorithm keeps the right endpoint $r$ of the last selected interval and the set $A$ of symbols already used. When interval $I=[\ell,u]$ associated with symbol $a$ is processed, the algorithm selects $I$ if $\ell>r$ and $a\notin A$; it then sets $r=u$ and inserts $a$ into $A$.

\begin{theorem}
\label{thm:greedy-two}
The greedy algorithm is a deterministic $2$-approximation for \MaxLL and runs in $O(n)$ time.
\end{theorem}

\begin{proof}
Let $G\subseteq\mathcal I(S)$ be the family of intervals selected by the greedy algorithm, and let $O\subseteq\mathcal I(S)$ be a maximum-cardinality feasible family of intervals representing an optimum \MaxLL solution, in consecutive-occurrence normal form. Consider an interval $I\in O$ when it is processed. If $I\notin G$, then either its symbol was already selected by the greedy algorithm or $I$ intersects the last greedy interval $J$.

In the first case, charge $I$ to the greedy interval having the same symbol. Since $O$ contains at most one interval per symbol, each greedy interval receives at most one charge of this type.

In the second case, the right endpoint of $J$ is at most the right endpoint of $I$ because intervals are processed by nondecreasing right endpoint. Since $I$ intersects $J$ and starts no later than the right endpoint of $J$, the interval $I$ contains the right endpoint of $J$. The intervals of $O$ are pairwise disjoint, so at most one interval of $O$ can contain that point. Hence every greedy interval receives at most one charge of the second type.
To make the latter bound explicit, fix a greedy interval $J$ with right endpoint $r_J$. Every optimum interval rejected on account of $J$ has left endpoint at most $r_J$ and right endpoint at least $r_J$, so it contains the same position $r_J$. Two distinct intervals of $O$ cannot both contain $r_J$, because the optimum family is pairwise disjoint. Thus this charging is injective for each fixed $J$.

Charge each interval of $O\cap G$ to itself. If $J\in O\cap G$, no other optimum interval has its symbol, so $J$ receives no charge of the first rejection type and at most one charge of the second type. If $J\notin O$, then $J$ receives at most one charge of each rejection type. Thus every greedy interval accounts for at most two intervals of $O$, and $|O|\le 2|G|$. The intervals are generated online in right-endpoint order by a single scan of $S$, so the running time is $O(n)$ with standard constant-time symbol dictionaries.
\end{proof}

\subsection{Equivalence with Job Interval Selection}

In \JISP, every job has a finite set of possible intervals on the real line. A feasible schedule chooses at most one interval for every job, and all chosen intervals must be pairwise non-overlapping. The objective is to maximize the number of scheduled jobs~\cite{ChuzhoyEtAl2006,ErlebachSpieksma2003}.

For every symbol $a$, create one job $J_a$ whose possible intervals are the intervals $I^a_i$ between consecutive occurrences of $a$.

\begin{theorem}
\label{thm:jisp-equivalence}
The optimum value of \MaxLL on $S$ equals the optimum value of the corresponding \JISP instance.
\end{theorem}

\begin{proof}
By Proposition~\ref{prop:interval-characterization}, every \MaxLL solution is a pairwise disjoint family containing at most one interval for each symbol, hence a feasible \JISP schedule of the same cardinality. Conversely, every feasible \JISP schedule selects pairwise disjoint intervals and at most one interval from each symbol-job, so Proposition~\ref{prop:interval-characterization} turns it into a \MaxLL solution of the same value.
\end{proof}

Chuzhoy, Ostrovsky, and Rabani~\cite{ChuzhoyEtAl2006} give, for every constant $\varepsilon>0$, a polynomial-time randomized $(e/(e-1)+\varepsilon)$-approximation for \JISP, with the guarantee holding in expectation. Therefore we obtain the following consequence.

\begin{corollary}
\label{cor:jisp-approx}
For every constant $\varepsilon>0$, \MaxLL admits a polynomial-time randomized $(e/(e-1)+\varepsilon)$-approximation algorithm.
\end{corollary}

\subsection{A $(4/3+\varepsilon)$-approximation for frequency four}

For a symbol occurring four times at positions $p_1<p_2<p_3<p_4$, write $I_1=[p_1,p_2]$, $I_2=[p_2,p_3]$, and $I_3=[p_3,p_4]$. The middle interval intersects both outer intervals, while $I_1$ and $I_3$ are disjoint.

We transform a $4$-\MaxLL instance to \JISPsub{2}. If a symbol occurs twice, create one singleton job containing its unique interval. If it occurs three times, create two singleton jobs, one for each consecutive interval; the two intervals overlap, so at most one of these jobs can be scheduled. If it occurs four times, create one job with alternatives $I_1$ and $I_3$, and one singleton job containing $I_2$.

\begin{lemma}
\label{lem:jisp2-reduction}
The transformation above preserves the optimum value.
\end{lemma}

\begin{proof}
For a two- or three-occurring symbol, feasibility is immediate. Consider a four-occurring symbol. A feasible \JISP schedule chooses at most one of $I_1,I_3$ because they belong to one job. It also cannot choose $I_2$ together with either $I_1$ or $I_3$ because the corresponding intervals overlap. Hence at most one interval associated with the original symbol is selected. Conversely, every feasible choice of one original candidate interval is available in the constructed instance. Applying this independently to every symbol gives a value-preserving correspondence between feasible solutions.
\end{proof}

Chuzhoy et al.~\cite{ChuzhoyEtAl2006} prove that their algorithm has approximation factor $k^k/(k^k-(k-1)^k)+\varepsilon$ for \JISPsub{k}; for $k=2$ this is $4/3+\varepsilon$.

\begin{theorem}
\label{thm:four-approx}
For every constant $\varepsilon>0$, $4$-\MaxLL admits a polynomial-time randomized $(4/3+\varepsilon)$-approximation algorithm.
\end{theorem}

\begin{proof}
By Lemma~\ref{lem:jisp2-reduction}, a $4$-\MaxLL instance reduces without loss to \JISPsub{2}. Apply the approximation algorithm of Chuzhoy et al.~\cite{ChuzhoyEtAl2006} and translate the resulting schedule back.
\end{proof}

\subsection{A complete frequency dichotomy}

\begin{theorem}
\label{thm:dichotomy}
For every fixed integer $d$, $d$-\MaxLL is solvable in linear time when $d\le 3$, and is APX-complete when $d\ge 4$.
\end{theorem}

\begin{proof}
For $d\le 3$, Theorem~\ref{thm:three-linear} applies. For every $d\ge 4$, APX-hardness follows because $4$-\MaxLL is a restriction of $d$-\MaxLL. Membership in APX follows from the deterministic $2$-approximation of Theorem~\ref{thm:greedy-two}.
\end{proof}

\section{An exact parameterized algorithm}
\label{sec:fpt}

Let $D=\{a\in\Sigma:\occ(a)>3\}$ and let $p=|D|$. We call the symbols of $D$ frequent. By Lemma~\ref{lem:consecutive}, it suffices to work with $\mathcal I(S)$. If $a\notin D$, then $a$ has at most two candidate intervals, and if it has two they overlap. Therefore pairwise disjointness alone guarantees that a feasible interval family contains at most one interval of every non-frequent symbol. Only the $p$ frequent symbols have to be represented explicitly in a dynamic-programming state. Such a state consists of an interval-prefix index $j$ and a subset $X\subseteq D$ of frequent symbols permitted to appear in the selected intervals; the subset is needed to prevent selecting the same frequent symbol more than once.

Let $\mathcal I(S)=\{I_1,\ldots,I_m\}$ be ordered by increasing right endpoint, where $I_j=[\ell_j,r_j]$ and $a_j$ is its symbol. Define $\pi(j)=\max(\{h<j:r_h<\ell_j\}\cup\{0\})$. The values $\pi(j)$ can be computed in linear time after the intervals have been generated in right-endpoint order.

For $j\in\{0,\ldots,m\}$ and $X\subseteq D$, let $F[j,X]$ be the maximum cardinality of a pairwise disjoint subfamily of $\{I_1,\ldots,I_j\}$ that contains at most one interval per symbol and uses no frequent symbol outside $X$. We set $F[0,X]=0$ for all $X$.

If $a_j\notin D$, then $F[j,X]=\max\{F[j-1,X],1+F[\pi(j),X]\}$. If $a_j\in D$ but $a_j\notin X$, then $F[j,X]=F[j-1,X]$. Finally, if $a_j\in X$, then
\begin{equation}
\label{eq:fpt-rec}
F[j,X]=\max\{F[j-1,X],\ 1+F[\pi(j),X\setminus\{a_j\}]\}.
\end{equation}

\begin{lemma}
\label{lem:fpt-correct}
For every $j$ and $X\subseteq D$, the value $F[j,X]$ equals the optimum cardinality among pairwise disjoint subfamilies of $\{I_1,\ldots,I_j\}$ that use at most one interval per symbol and no frequent symbol outside $X$.
\end{lemma}

\begin{proof}
The proof is by induction on $j$. An optimum family either excludes $I_j$, yielding the first term, or includes $I_j$, in which case every other chosen interval belongs to $\{I_1,\ldots,I_{\pi(j)}\}$. For a non-frequent symbol, no additional state is needed because all of its candidate intervals intersect one another. For a frequent symbol $a_j$, selecting $I_j$ forbids selecting another interval of $a_j$ in the prefix, which is exactly enforced by replacing $X$ with $X\setminus\{a_j\}$. These are the only possibilities.
\end{proof}

\begin{theorem}
\label{thm:fpt}
\MaxLL can be solved exactly in $O(2^p n)$ time, where $p=|\{a\in\Sigma:\occ(a)>3\}|$.
\end{theorem}

\begin{proof}
The answer is $F[m,D]$. There are $m\le n-1$ candidate intervals and $2^p$ subsets of $D$, and each table entry is computed in constant time from previously computed entries. Candidate intervals and predecessor indices can be generated in $O(n)$ time, so the total running time is $O(2^p n)$.
\end{proof}

The parameter $p$ measures the distance from the tractable frequency-three regime: $p=0$ is exactly the setting in which the interval-color constraint disappears.

\section{Single-deletion sensitivity}
\label{sec:sensitivity}

A standard question for a string optimization problem is how much its optimum changes after a small modification of the input. Such \emph{sensitivity} questions arise, for example, when symbols represent uncertain observations or sequencing errors: one would like to know whether the optimum depends critically on a particular position without solving the problem independently for each modified string. We consider three elementary edit operations: a \emph{deletion} removes one symbol at a specified position, an \emph{insertion} adds a new symbol at a specified boundary, and a \emph{substitution} replaces the symbol at one position by another symbol. In the remainder of this section the main algorithmic task is \emph{offline single-deletion sensitivity}: independently delete each of the $n$ original positions and report the resulting optimum. The edits do not accumulate; every query is applied to the same original string $S$.

Let $\mu(S)$ denote the optimum value of \MaxLL on $S$.

\begin{lemma}[Edit stability]
\label{lem:lipschitz}
If $S'$ is obtained from $S$ by one deletion, then $\mu(S)-1\le\mu(S')\le\mu(S)$. If $S'$ is obtained by one insertion, then $\mu(S)\le\mu(S')\le\mu(S)+1$. If $S'$ is obtained by one substitution, then $|\mu(S')-\mu(S)|\le 1$.
\end{lemma}

\begin{proof}
Deletion cannot create a subsequence that was unavailable before, so $\mu(S')\le\mu(S)$. Take an optimum solution of $S$. If the deleted position is not used, the same solution remains feasible. Otherwise it belongs to one represented symbol $a$. Removing the complete $a$-run, if necessary, destroys at most that one represented symbol and leaves a feasible letter-duplicated subsequence, so $\mu(S')\ge\mu(S)-1$. The insertion inequalities are the same statement read in reverse. A substitution can be viewed as a deletion followed by an insertion, but both strings share the string obtained by deleting the substituted position; applying the deletion inequality to both gives the absolute difference bound of one.
\end{proof}

We now compute all deletion answers simultaneously. Fix $D=\{a:\occ_S(a)>3\}$ and $p=|D|$. Tracking a symbol in $D$ remains safe even if, after a deletion, its frequency drops to three.

For a boundary $i$ and $A\subseteq D$, let $L[i,A]$ be the optimum number of intervals contained in $S[1..i]$ whose set of selected frequent symbols is exactly $A$. Symmetrically, let $R[i,B]$ be the optimum number of intervals contained in $S[i..n]$ that use no frequent symbol outside $B$. Both tables can be computed in $O(2^p n)$ time by left-to-right and right-to-left variants of the dynamic program of Section~\ref{sec:fpt}.

For a candidate interval $I=[u,v]$ associated with symbol $a$, write $F_a=\{a\}$ if $a\in D$ and $F_a=\emptyset$ otherwise. Define its forced score by
\begin{equation}
\label{eq:forced-score}
W(I)=1+\max_{A\subseteq D\setminus F_a}\bigl(L[u-1,A]+R[v+1,D\setminus(A\cup F_a)]\bigr).
\end{equation}

\begin{lemma}
\label{lem:forced-score}
The value $W(I)$ is the optimum value among solutions forced to contain $I$.
\end{lemma}

\begin{proof}
Once $I$ is selected, every other interval lies entirely to its left or entirely to its right. Frequent symbols used on the two sides must be disjoint and must exclude $a$ when $a\in D$, which is enforced by the partition $A$ and $D\setminus(A\cup F_a)$. A non-frequent symbol cannot be selected on both sides: doing so would require two disjoint consecutive-occurrence intervals of that symbol, hence at least four occurrences, contradicting that it is non-frequent. Thus the two side solutions combine without hidden symbol conflicts, and maximizing over $A$ gives exactly the optimum forced score.
\end{proof}

For a position $x$, define
\begin{equation}
\label{eq:no-cross}
N_x=\max_{A\subseteq D}\bigl(L[x-1,A]+R[x+1,D\setminus A]\bigr).
\end{equation}
This is the best solution after deleting $x$ that uses no interval crossing the deleted position.

Deleting $x$ changes consecutive-occurrence intervals only for the symbol $s_x$. If $u<x<v$ are respectively the previous and next occurrences of $s_x$, then deletion creates the bridge interval $[u,v]$; if one of them does not exist, no bridge is created. Every old candidate interval with $x$ in its interior survives the deletion, while candidate intervals having $x$ as an endpoint disappear.

\begin{theorem}
\label{thm:all-deletions}
Let $p=|\{a:\occ_S(a)>3\}|$. The values $\mu(S-x)$ for all positions $x\in\{1,\ldots,n\}$ can be computed in $O(2^p n+n\log n)$ total time and $O(2^p n)$ space.
\end{theorem}

\begin{proof}
Fix $x$. Any feasible interval family after deleting $x$ falls into one of three classes. First, no selected interval crosses the deleted position; the optimum of this class is $N_x$. Second, one surviving old interval $I=[u,v]$ satisfies $u<x<v$. There can be at most one such selected interval because selected intervals are pairwise disjoint, and the optimum conditional on selecting it is $W(I)$. Third, the selected crossing interval is the new bridge between the previous and next occurrences of $s_x$; its optimum value is given by the same expression as in~\eqref{eq:forced-score}, using the bridge endpoints and excluding its symbol when frequent. These cases are exhaustive because deletion changes the occurrence adjacency only for $s_x$.

Hence $\mu(S-x)$ is the maximum of $N_x$, the forced scores of old candidate intervals whose interiors contain $x$, and the forced score of the bridge interval when it exists. The prefix and suffix tables require $O(2^p n)$ time. Computing $N_x$ for all $x$, all old forced scores, and all bridge scores also takes $O(2^p n)$ time because there are $O(n)$ relevant positions and intervals. It remains to compute, for every $x$, the maximum $W(I)$ over old intervals $I=[u,v]$ with $u<x<v$. This is an offline interval range-maximum problem: interval $I$ contributes value $W(I)$ to positions $u+1,\ldots,v-1$. A segment tree computes all these maxima in $O(n\log n)$ total time. Combining the three values yields every deletion answer.
\end{proof}

For $3$-\MaxLL, we have $p=0$, and Theorem~\ref{thm:all-deletions} gives all $n$ single-deletion answers in $O(n\log n)$ total time.

\section{Longest Run Subsequence}
\label{sec:lrs}

We next return to the Longest Run Subsequence problem (\LRS), formally defined in Section~\ref{sec:framework} and introduced by Schrinner et al.~\cite{SchrinnerEtAl2020,SchrinnerEtAl2021} in the context of homology-based genome scaffolding. Although \LRS maximizes the length of the chosen subsequence and \MaxLL maximizes the number of distinct represented symbols, both require each selected symbol to occupy a single contiguous block in the subsequence. 
Consequently, both can be modeled by non-overlapping occurrence intervals: in \MaxLL, an interval between consecutive equal occurrences has unit profit; in \LRS, a longer interval spanning the first and last occurrences of a selected run has profit equal to the number of occurrences in that run. This explains the use of weighted Job Interval Selection in the present section. 
We recall that the \LRS problem is NP-hard, and it remains APX-hard even when every symbol occurs at most twice~\cite{DondiSikora2021}.
For bounded-occurrence instances, Asahiro et al.~\cite{AsahiroEtAl2023} gave a polynomial-time $(k+1)/2$-approximation when every symbol occurs at most $k$ times, together with a $4/3$-approximation for $k=2$. More recently, Asahiro et al.~\cite{AsahiroEtAl2025} improved the ratios to $(k+2)/3$ for $k\in\{2,3\}$ and $2(k+1)/5$ for $k\ge4$. All these guarantees depend on $k$, and therefore they do not give a constant approximation ratio for unrestricted instances. We now remove this dependence by an exact reduction to one-machine weighted Job Interval Selection.

Let the occurrences of symbol $a$ be
$p^a_1<\cdots<p^a_{t_a}$. Create one weighted job $J_a$.
For every $1\le i\le j\le t_a$, create the half-open interval
$I^a_{i,j}=[2p^a_i,2p^a_j+1)$ of weight $j-i+1$.
This interval represents selecting all occurrences
$p^a_i,p^a_{i+1},\ldots,p^a_j$ as the unique run of $a$.

The transformation from a closed interval $[u,v]$ to the
half-open interval $[2u,2v+1)$ ensures that every interval
has positive length, including those corresponding to
runs of length one ($u=v$). Moreover, multiplying the
positions by two and adding one to the right endpoint
ensures that left and right endpoints never coincide.
The transformation preserves the intersection relation:
two closed intervals $[u,v]$ and $[u',v']$, with $u\le u'$,
are disjoint if and only if $v<u'$, which is equivalent
to $2v+1\le 2u'$ since the positions are integers.
Thus, two occurrence intervals are disjoint if and only
if their corresponding half-open intervals are disjoint.

Every pair of occurrence indices contributes one interval,
so the number of weighted intervals is
$M=\sum_{a\in\Sigma}t_a(t_a+1)/2\le n(n+1)/2$.
All jobs and intervals can therefore be generated in
$O(n^2)$ time and space.

\begin{algorithm}[t]
\caption{A $2$-approximation for Longest Run Subsequence}\label{alg:lrs-two}
\begin{algorithmic}[1]
\Require String $S=s_1\cdots s_n$
\Ensure A run subsequence of $S$
\State Compute the occurrence lists $P_a=(p^a_1,\ldots,p^a_{t_a})$ for all $a\in\Sigma$.
\ForAll{$a\in\Sigma$}
  \State Create a weighted job $J_a$.
  \For{$i=1,\ldots,t_a$}
    \For{$j=i,\ldots,t_a$}
      \State Add $I^a_{i,j}=[2p^a_i,2p^a_j+1)$ to $J_a$ with weight $j-i+1$.
    \EndFor
  \EndFor
\EndFor
\State Apply a factor-$2$ approximation algorithm for one-machine \WJISP.
\State Sort the returned intervals by left endpoint.
\State \Return the concatenation, in this order, of the runs represented by these intervals.
\end{algorithmic}
\end{algorithm}

\begin{lemma}
\label{lem:lrs-wjisp}
The optimum value of \LRS equals the optimum value of the constructed one-machine \WJISP instance.
\end{lemma}

\begin{proof}
Consider a feasible run subsequence. For every selected symbol $a$, let $p^a_i$ and $p^a_j$ be its first and last selected occurrences. Every occurrence of $a$ between these two positions can be added without creating another $a$-run: if a selected occurrence of another symbol lay between two selected $a$'s, the two $a$'s would already belong to different runs. Hence the run can be assumed to contain all occurrences $p^a_i,\ldots,p^a_j$ and has length $j-i+1$. Distinct runs occur in disjoint position ranges, so the corresponding half-open intervals are non-overlapping, and at most one interval is selected from each job.

Conversely, order the intervals of a feasible \WJISP solution from left to right and, for every selected $[2p^a_i,2p^a_j+1)$, take the occurrences $p^a_i,\ldots,p^a_j$. Non-overlap guarantees that the resulting blocks occur consecutively in the subsequence, and the one-interval-per-job constraint guarantees at most one run for each symbol. The total selected length equals the total interval weight.
\end{proof}

Erlebach and Spieksma~\cite{ErlebachSpieksma2003} observe that known LP-based and combinatorial algorithms yield a factor-$2$ approximation for one-machine weighted Job Interval Selection; see also~\cite{BarNoyEtAl2001,BermanDasGupta2000}.

\begin{theorem}
\label{thm:lrs-two}
\LRS admits a polynomial-time $2$-approximation algorithm without any bound on symbol frequencies.
\end{theorem}

\begin{proof}
Algorithm~\ref{alg:lrs-two} constructs the weighted-job instance in $O(n^2)$ time and space. By Lemma~\ref{lem:lrs-wjisp}, the instance has the same optimum value as the original \LRS instance. A factor-$2$ approximation for \WJISP returns a schedule of weight at least half that optimum; translating each selected interval into the corresponding consecutive block of occurrences preserves the weight and produces a valid run subsequence. Thus the returned subsequence has length at least $\OPT_{\LRS}(S)/2$. If the chosen \WJISP algorithm runs in time $T(M)$ on $M$ intervals, the overall running time is $O(n^2+T(M))$ with $M=O(n^2)$, and is polynomial.
\end{proof}

Theorem~\ref{thm:lrs-two} is independent of the maximum number of occurrences of any symbol. Thus it complements the stronger ratios known for small fixed $k$~\cite{AsahiroEtAl2023,AsahiroEtAl2025} while remaining constant as $k$ grows. To the best of our knowledge, it is the first constant-factor approximation for unrestricted \LRS.

\section{Conclusions and future work}
\label{sec:conclusion}

We developed an interval-based view of \MaxLL that unifies several aspects of the problem. The consecutive-occurrence normal form reduces the relevant candidate family to linear size. It leads to a simple deterministic $2$-approximation, connects the problem exactly to Job Interval Selection, gives a randomized $(4/3+\varepsilon)$ approximation for the frequency-four case, and yields a complete frequency dichotomy between the linear-time regime $d\le 3$ and the APX-complete regime $d\ge 4$.

The same viewpoint supports an exact $O(2^p n)$ algorithm parameterized by the number $p$ of symbols occurring more than three times. For the tractable frequency-three case, the greedy dynamics can be represented as a weighted-ancestor problem, giving fast substring queries. The prefix--suffix decomposition used for the parameterized dynamic program also leads to an offline sensitivity algorithm computing all independent single-deletion answers in $O(2^p n+n\log n)$ total time.

We also presented an exact connection from \LRS to weighted Job Interval Selection, giving a polynomial-time factor-$2$ approximation for unrestricted Longest Run Subsequence. Unlike the existing occurrence-dependent ratios, this factor is independent of symbol frequencies.

Several questions remain open. The first is whether the special structure of frequency-four instances permits a simple deterministic approximation ratio strictly below $2$, ideally approaching the randomized $4/3+\varepsilon$ guarantee. A second direction is to improve the dependence on $p$ in the exact parameterized algorithm, or to identify more refined structural parameters. It would also be interesting to obtain fully dynamic data structures under substitutions, insertions, and deletions, and to sharpen the preprocessing bounds for the range-query data structure by exploiting structural properties of the particular parent trees generated by \MaxLL instances.

\section*{AI disclosure.} The authors used GPT-5.6 Sol during the preparation of this
work to brainstorm and refine constructions, explore proof
strategies, and discuss and check mathematical arguments.
The model also assisted with proofreading and aspects of
the literature review. All mathematical claims, proofs,
and cited results were independently verified by the authors,
who take full responsibility for the accuracy and content
of the paper.

\bibliographystyle{elsarticle-num}
\bibliography{ref}

\end{document}